\documentstyle[aps]{revtex}            
\begin{document}


\draft
\preprint{}
\title{Quaternionic Electron Theory:\\
       Geometry, Algebra and Dirac's Spinors}
\author{Stefano De Leo\thanks{{\em E-Mail}: 
              {\tt deleos@le.infn.it} ,    
              {\tt deleo@ime.unicamp.br}    }$^{a,b}$ and   
        Waldyr A.~Rodrigues, Jr.\thanks{{\em E-Mail}:
              {\tt walrod@ime.unicamp.br}    }$^{b}$}
\address{$^{a}$Dipartimento di Fisica, 
                Universit\`a degli Studi Lecce and INFN, 
                Sezione di Lecce\\
                via Arnesano, CP 193, 73100 Lecce, Italia\\
                and\\ 
         $^{b}$Instituto de Matem\'atica, Estat\'{\i}stica e 
                Computa\c{c}\~ao Cient\'{\i}fica, IMECC-UNICAMP\\
                CP 6065, 13081-970, Campinas, S.P., Brasil} 
\date{October, 1997}
\maketitle


\begin{abstract}

The use of complexified quaternions and $i$-complex geometry 
in formulating the Dirac equation allows
us to give {\em interesting} geometric interpretations hidden in the 
conventional matrix-based approach. 

\end{abstract}


\section{Introduction}
\label{s1}

Among the many alternative mathematical systems in which the Dirac equation
can be written~\cite{ZUB,HES,RQUA,KEL,RQUA1,RQUA2,CQUA}, we showed in a recent 
paper~\cite{DR} how the matrix and vector algebras can be replaced by a 
single mathematical system, complexified quaternionic algebra, with which 
geometric interpretations can be carried out more efficiently. The power of our
(complexified) quaternionic formalism becomes evident within the electron 
theory, and it derives from the fact that the elements of the complexified 
quaternionic algebra are subject to direct geometrical identifications.
The theory presented in our previous paper~\cite{DR} is algebraically 
isomorphic
to the matrix Dirac theory; it can be provided with an equivalent physical
interpretation as well. It differs from the Dirac approach in that all its
algebraic ingredients have a geometrical significance determined by
geometric properties of the complexified quaternionic algebra. 
In contrast to the
complex matrix algebra of Dirac, the complexified quaternionic algebra has
a clear geometrical significance. By introducing new imaginary units,
we give a geometrical interpretation for the ``complex'' imaginary unit, $i$,
which characterizes standard quantum theories. 

By working with complexified quaternions and $i$-complex 
geometry~\cite{DR,DR1}, we also obtain the surprising conclusion that the 
Schr\"odinger theory contains additional informations.  
It is possible to affirm that spin and positron are 
already present in the complexified quaternionic Schr\"odinger theory. At first
sight such a conclusion may seem preposterous. Schr\"odinger knew nothing
of spin and positron when he framed his equation. ``But it is no more
preposterous than the incredible fact that Schr\"odinger wrote down his 
equation and solved the first problems of modern quantum theory without any 
mention of probability, thought it now appears that probability was 
already in the theory. The Born 
interpretation of $\psi^{\dag} \psi$ as a probability density 
can be adopted without the
slights modification of the Schr\"odinger work. Indeed, the Born  
interpretation is now an indispensable part of the theory''~\cite{HG}.

Let us give an {\em informal} discussion concerning the additional solutions 
in the complexified quaternionic Schr\"odinger equation. By working with
complex numbers we have no possibility to accommodate the two freedom degrees
characterizing the spin of the electron
\begin{equation}
\label{cs} 
\psi \in C(1,i)~~~~~~~[~\mbox{electron with ``frozen'' spin}~]~.
\end{equation}
By allowing real quaternions as underlying numerical field we double the
complex freedom degrees, and consequently we have rotations in the plane
\[ \left(~C_{(1,i)}~,~jC_{(1,i)}~\right)~.\]
In this case we can accommodate the spin up and down of the electron 
respectively on the complex and {\em pure} quaternionic axis. Thus, (real)
quaternionic wave-functions naturally describe spin in the Schr\"odinger theory,
\begin{equation}
\psi \in {\cal H}~~~~~~~[~\mbox{electron with spin up and down}~]~.
\end{equation}
Finally, by considering complexified quaternionic numbers we can introduce
the dual space $\bbox{\iota}{\cal H}$ and quadruple the complex degrees of
freedom
\[ \left(~{\cal H}~,~\bbox{\iota}{\cal H}~\right)~.\]
This means that the wave-function, $\Psi$, in ${\cal H}_c$,  possesses the 
needed degrees of freedom to represent the spin and positron content
\begin{equation}
\Psi \in {\cal H}_c~~~~~~~[~\mbox{electron/positron with spin up and down}~]~.
\end{equation}
Obviously, the previous discussion represents only a ``rough'' approach. 
In section~\ref{s3}, 
by analyzing the complexified quaternionic Dirac spinors, 
we shall give a more clear geometric interpretation. Positive and 
negative energy solutions are respectively characterized (in the rest frame) 
by real and pure complexified quaternionic spinors. For the most general 
solutions, we will be able to related them to 
rotation angles in the complexified quaternionic hyper-plane
\[ \left(~{\cal H}_c~,~\bbox{\iota}{\cal H}_c~\right)~.\]
Finally, the {\em polar form} of complexified quaternionic spinors, 
given in section~\ref{s2}, will allow to identify the spin operator by a 
{\em new} hypercomplex imaginary unit $\cal I$, individued by the  
components of the momentum operator $\vec{p}$ of our particle.


\section{Geometric Algebra of Complexified Quaternions}
\label{s2}

In this section we draw out the {\em polar form} of complexified quaternions,
which will be very useful in discussing geometric interpretations 
of Dirac spinors.

Let us start with the standard discussion about plane geometry and 
complex algebra. We may relate the geometry to the complex algebra by 
representing complex numbers in a plane
\begin{equation} 
\label{fe}
x+i y = r e^{i \theta}~.
\end{equation}
In fact, we know that a rotation of $\alpha$-angle around the $z$ axis, can be 
represented by $e^{i\alpha}$, 
\[ e^{i \alpha} \left( x+i y \right) = r e^{i (\alpha + \theta)}~.\]
Feynman called Eq.~(\ref{fe}), the unification of algebra and 
geometry~\cite{FEY}. We like talking of ``partial'' unification. Our aim in 
this section is to show how it is possible generalize the connection between 
geometry and algebra by using a noncommutative numerical field, or more
generally hypercomplex numbers.

Similarly to rotations in a plane, a rotation about an axis passing 
through the origin and parallel to a given unitary vector 
$\hat{u}\equiv (u_x, u_y, u_z)$ by an angle $\alpha$ can be obtained 
taking the real quaternionic transformation
\begin{equation}
\exp \left( \frac{iu_x+ju_j+ku_z}{2} \alpha \right)
\, \left(ix+jy+kz \right)
\, \exp \left( -\frac{iu_x+ju_j+ku_z}{2} \alpha \right)~.
\end{equation}
The vector part ($q^{\dag}=-q$) of a generic quaternion
\[ q = r_0 + \vec{h} \cdot \vec{r} ~ , \]
with the real axis $q=r_0$ specifies a unique ``complex'' plane with imaginary
axis given by the unit
\[ i \, \frac{r_1}{r} + j \, \frac{r_2}{r} + k \, 
\frac{r_3}{r}~,~~~~~~~ 
   r=\sqrt{r_1^2+r_2^2+r_3^2}~.\]
Our quaternionic number will be represented in this plane by
\begin{equation}
 r_0 + I   r = \rho e^{I \alpha}~,
\end{equation}
where
\[ 
I = \frac{\vec{h} \cdot \vec{r}}{r}~,~~~~~\vec{h} \equiv (i,j,k)~.
\]

Let us analyze what happens for complexified quaternions. 
What is the {\em polar form} of complexified quaternions? How can we 
individuate a ``complex'' phase? It is possible to define an hyper-plane?

A generic complexified quaternion is expressed by
\begin{equation}
q_c = c_0 + i c_1 + j c_2 + k c_3~,~~~~~~~c_{0,1,2,3} 
\in {\cal C}(1,\bbox{\iota})~.
\end{equation}
We define
\[ c = \sqrt{c_1^2+c_2^2+c_3^2}~,\]
and consequently we have
\begin{equation}
q_c = c_0 + {\cal I} c~,
\end{equation}
where
\[ {\cal I} = i \, \frac{c_1}{c} + j \, \frac{c_2}{c} + k \, 
\frac{c_3}{c} = \frac{\vec{h} \cdot \vec{c}}{c}
~,~~~~~~~{\cal I}^2=-1~,~~~~~~~{\cal I}^{\star} {\cal I} =1~.\] 
The vector part ($q_c^{\star}=-q_c$) of a generic complexified quaternion 
$q_c$ with the $\bbox{\iota}$-complex axis $q_c=c_0$ specifies a unique 
{\em hypercomplex} plane with ``imaginary'' axis given by the unit $\cal I$.

By taking the $\star$-involution, we find
\[ q_c^{\star} q_c = c_0^2 + c ^2 \in {\cal C}(1,\bbox{\iota})~.\]
Thus, we can write
\[ q_c^{\star} q_c = \rho^2 e^{2\bbox{\iota} \alpha}~~~~~~~~
\rho, \alpha \in {\cal R}~,\]
and consequently to identify
\[ c_0 = \rho e^{\bbox{\iota} \alpha} \cos z~,~~~~~ 
   c= \rho e^{\bbox{\iota} \alpha} \sin z~~~~~~~ 
z\in {\cal C}(1,\bbox{\iota})~.\] 
Then, the {\em polar form} of a complexified quaternion reads
\begin{equation}
q_c =  \rho e^{\bbox{\iota} \alpha} e^{{\cal I} z}~,
\end{equation} 
where
\[\rho  e^{\bbox{\iota} \alpha} =  \sqrt{c_0^2+c_1^2+c_2^2+c_3^2}~, \]
and
\[ \cos z = \frac{c_0}{\sqrt{c_0^2+c_1^2+c_2^2+c_3^2}}~,~~~~~~~
   \sin z = \frac{\sqrt{c_1^2+c_2^2+c_3^2}}{\sqrt{c_0^2+c_1^2+c_2^2+c_3^2}}~.
\]

In conclusion we can characterize a complexified quaternionic number by 
{\em two} ``imaginary'' units
\[ \bbox{\iota}~~~~~~~\mbox{and}~~~~~~~{\cal I}~,\]
respectively related to the following ``hypercomplex'' planes
\[ \Lambda^{(E)} \equiv 
\left( \, \rho e^{{\cal I}z} \cos \alpha  \, , \,
          \rho e^{{\cal I}z} \sin \alpha  \, \right)~~~~~~~\mbox{and}~~~~~~~
 \Lambda^{(s)} \equiv 
\left( \, \rho e^{\bbox{\iota} \alpha} \cos z  \, , \,
          \rho e^{\bbox{\iota} \alpha} \sin z \, \right)~.
\]
The first plane will identify the mixing of positive and negative energy
solutions, whereas the second one will represent the spin plane.
                         
The real quaternionic theory essentially distinguishes two types of objects,
the scalars and the vectors. Nevertheless, we know that the vectors split 
into two disjoint sets, polar and axial vectors (often called pseudo-vectors) 
and, in three-dimensions we have also scalars and pseudo-scalars. These 
distinctions are well illustrated by complexified quaternions~\cite{LIB1}. 
Following the convention used by Hestenes~\cite{LIB2}, the terms in
complexified quaternions
\[         \alpha_0 + \bbox{\iota} \beta_0
         + i ( \alpha_1 + \bbox{\iota} \beta_1 )    
         + j ( \alpha_2 + \bbox{\iota} \beta_2 )    
         + k ( \alpha_3 + \bbox{\iota} \beta_3 )~~~~~~~
\alpha_{0,1,2,3}, \, \beta_{0,1,2,3} \in {\cal R}~, 
\]
naturally separate into four groups
\begin{eqnarray*}
\alpha & ~~~\rightarrow~~~& \mbox{scalars}~,\\
\bbox{\iota} \beta & ~~~\rightarrow~~~& \mbox{pseudo-scalars}~,\\
\bbox{\iota} \vec{h} \cdot \vec{\beta} & ~~~\rightarrow~~~& \mbox{vectors}~,\\
\vec{h} \cdot \vec{\alpha} & ~~~\rightarrow~~~& \mbox{bivectors}~.
\end{eqnarray*}
In the complexified quaternionic electron theory we can define the dual of 
a complexified quaternion, $q_c$, to be $\bbox{\iota} q_c$. The dual operation
turns a scalar into a pseudo-scalar and a vector into a bivector
(and vice-versa). We showed~\cite{DR} that parity operation, in the Dirac 
theory, is
related to the $\bullet$-involution. Thus, in the present formalism obtaining
the dual is simply achieved by multiplying by $\bbox{\iota}$.


\section{Quaternionic Dirac's Spinors}
\label{s3}

The possibility to express the Dirac spinor as a simple complexified 
quaternions instead of vector column (as in real quaternionic and complex 
formulations) represents the main difference between complexified 
quaternionic and real quaternionic or complex version of the Dirac equation. 

``The  
representation of a Dirac spinor by a generic (invertible) element of the 
Pauli algebra has given rise to new insights in Dirac's theory of the 
electron: the polar form of Dirac-Hestenes spinor presents the Dirac spinor
as the product of a {\em complex} number by a Lorentz transformation, where 
the argument of the complex number has been identified with the 
Yvone-Takabayashi angle and its module with the charge density of the 
mixture of positrons and electrons. Nevertheless the proof of the equivalence 
between the original Dirac equation and the Dirac-Hestenes equation is 
usually made raising an explicit relation between the components of Dirac 
spinor and the Dirac-Hestenes spinor, that require a lot of calculations to 
be done'' - {\em Zeni}~\cite{ZEN}. 

Our complexified quaternionic approach to Dirac theory 
reproduce quickly the standard results, avoids a lot of calculations and 
gives the desired {\em geometrical interpretations} which characterize the 
Dirac-Hestenes theory~\cite{LIB2}. The plane wave solutions of the 
complexified quaternionic Dirac equation 
\begin{equation}
\label{de}
\left( \partial_t  +\bbox{\iota} \vec{h} \cdot \vec{\partial} \right) \Psi(x)
\, i = m \Psi^{\bullet}(x)~,
\end{equation}
are
\begin{center}
$\sqrt{\frac{|E|+m}{2}}$~$\times$~
\begin{tabular}{|llc|}\hline \hline
 & & \\
~~$1 + \frac{\bbox{\iota} \vec{h} \cdot \vec{p}}{|E|+m}$~, &
~~~$\left( 1 + \frac{\bbox{\iota} \vec{h} \cdot \vec{p}}{|E|+m} \right) \, j$
~,
 & ~~~$E>0$~~~\\
 & & \\
\hline
 & & \\
~$\left( 1 - \frac{\bbox{\iota} \vec{h} \cdot \vec{p}}{|E|+m} \right) \, 
\bbox{\iota}$~, &
~~~$\left( 1 - \frac{\bbox{\iota} \vec{h} \cdot \vec{p}}{|E|+m} \right) \, 
\bbox{\iota} j$~,
 & ~~~$E<0$~~~\\
 & & \\
\hline \hline
\end{tabular}
~$\times$~$e^{-ipx}$~.
\end{center}

The wave-function does not have a direct physical significance, and a 
crucial part of the Dirac theory is to relate $\Psi$ to observable 
quantities. The {\em polar} decomposition of $\Psi$
\begin{equation}
\label{1}
\rho \, \exp (\bbox{\iota} \alpha) \, \exp ({\cal I}z)
\end{equation}
greatly facilitates this 
task and, in addition, makes the geometric content of the theory explicit. The
quantities 
\[ \rho,~\alpha,~{\cal I},~z~\]
have distinctive geometrical and physical interpretations which are 
independent of any matrix representation. So, it is best to use them 
instead of the $\alpha$'s and $\beta$'s which appear in the standard Dirac 
equation.

Ordinarily, Dirac spinors are said to be representations of the Lorentz group
because they transform in a certain way under Lorentz transformation. In 
contrast, we say that $\Psi$ represents a Lorentz transformation, the spinor
``$e^{{\cal I}z}$'' 
may be regarded as a representation of a Lorentz transformation. 
Complexified quaternionic spinors, $\Psi$, consist of Lorentz rotations, 
dilatation and duality transformation. 

By looking at the Dirac plane wave solutions in the rest frame of the particle
($\vec{p}=\vec{0}$), we immediately find the following structures for our 
spinors
\begin{eqnarray*}
\Psi_{E=+m} \in {\cal H} 
  & ~~~\rightarrow~~~ & \alpha = 0~,\\ 
\Psi_{E=-m} \in \bbox{\iota} {\cal H} 
  & ~~~\rightarrow~~~ & \alpha = \frac{\pi}{2}~.
\end{eqnarray*}
Thus, the hyper-plane $\left( {\cal H}, \bbox{\iota} {\cal H} \right)$, gives 
the mixing of positive/negative energy solutions in the rest frame of the
particle. This was anticipated in our introduction. 
The situations appears more complicated for the general case in which 
the particle is in motion. In such a case the positive energy solutions are not
identified by simple (real) quaternions and so it is not immediate to extract
$\alpha =0$. Nevertheless, by using the $\star$-conjugation operation
($\vec{h} \rightarrow -\vec{h}$, $\bbox{\iota} \rightarrow \bbox{\iota}$)
we find
\[
\Psi^{\star} \Psi =  
\rho^2 \, \exp (2\bbox{\iota} \alpha)~, 
\]
and consequently the Eqs.
\begin{eqnarray*}
\Psi_{E>0}^{\star} \Psi_{E>0} & = &
\left(1 - \frac{\bbox{\iota} \vec{h} \cdot \vec{p}}{|E|+m} \right)
\left(1 + \frac{\bbox{\iota} \vec{h} \cdot \vec{p}}{|E|+m} \right) =
-j\left(1 - \frac{\bbox{\iota} \vec{h} \cdot \vec{p}}{|E|+m} \right)
\left(1 + \frac{\bbox{\iota} \vec{h} \cdot \vec{p}}{|E|+m} \right) j=
\frac{2m}{|E|+m}~,\\
\Psi_{E<0}^{\star} \Psi_{E<0} & = &
\bbox{\iota}^2
\left(1 + \frac{\bbox{\iota} \vec{h} \cdot \vec{p}}{|E|+m} \right)
\left(1 - \frac{\bbox{\iota} \vec{h} \cdot \vec{p}}{|E|+m} \right) =
-\bbox{\iota}^2 j
\left(1 - \frac{\bbox{\iota} \vec{h} \cdot \vec{p}}{|E|+m} \right)
\left(1 + \frac{\bbox{\iota} \vec{h} \cdot \vec{p}}{|E|+m} \right) j=
-\frac{2m}{|E|+m}~,
\end{eqnarray*}
imply that positive energy solutions are characterized by the
$\alpha$-angle value $0$, whereas negative energy solutions by the 
$\alpha$-angle value $\frac{\pi}{2}$. This means that we can think to 
an ``complex'' hyper-plane
\[ 
\left( 
\, {\cal H}_c^{\alpha=0} \, , \,  \bbox{\iota} {\cal H}_c^{\alpha=0} \,
\right)~\equiv~
\left( 
\, {\cal H}_c^{\alpha=0} \, , \,  {\cal H}_c^{\alpha=\frac{\pi}{2}} \,
\right)
~, \]
where the positive energy solutions are mapped in the ``real'' axis, whereas
the negative energy solutions are obtained by rotations of $\frac{\pi}{2}$
angles.

Let us now give the explicit {\em polar} form for our 
complexified quaternionic 
Dirac spinors. Consider
\[ 
\sqrt{\frac{|E|+m}{2}}~\times~
\left( 1 + \frac{\bbox{\iota} \vec{h} \cdot \vec{p}}{|E|+m}\right)~.
\]
It is immediate to recognize
\[ c_0 = \sqrt{\frac{|E|+m}{2}}~,~~~~~
   \vec{c} = \bbox{\iota} \frac{\vec{p}}{\sqrt{2(|E|+m)}}~,
\]
thus
\[ 
\rho e^{\bbox{\iota} \alpha} = 
\sqrt{\frac{|E|+m}{2}-\frac{\vec{p}^{\, 2}}{2(|E|+m)}} = \sqrt{m}~,
\]
and
\[ \cos z = \sqrt{\frac{|E|+m}{2m}}~,~~~~~
   \sin z = \bbox{\iota} \frac{|\vec{p}|}{\sqrt{2m(|E|+m)}}~.
\]
The ``generalized'' imaginary unit $\cal I$ is identified by 
\[ {\cal I} = \frac{\vec{h} \cdot \vec{p}}{|\vec{p}|}~. \]
We have now all the needed ingredients to write down the {\em polar form}
for our complexified quaternionic Dirac spinor
\[
\sqrt{\frac{|E|+m}{2}}~\times~
\left( 1 + \frac{\bbox{\iota} \vec{h} \cdot \vec{p}}{|E|+m}\right)
~~~\rightarrow~~~
\sqrt{m} \, e^{\bbox{\iota} {\cal I} \beta}~,
\]
where
\[ \cosh \beta = \sqrt{\frac{|E|+m}{2m}}~.\]
We can recognize in the definition of the ``generalized'' imaginary 
unit $\cal I$, the helicity operator~\cite{DR}. 
For $\vec{p}\equiv (p_x,0,0)$, it reduces 
to the complex imaginary unit $i$, characterizing the standard Quantum 
Mechanics. Thus, the $i$ in the Dirac equation can be interpreted
geometrically (bivector) and its appearance is strictly related to the
particle spin content. The {\em polar form} for the remaining Dirac spinors
is very simply 
\begin{center}
\begin{tabular}{ll}
$\sqrt{m} \, e^{\bbox{\iota} i \beta}~,$ & 
~~~~~$E>0~,~~\mbox{spin}~ \uparrow~,$\\
$\sqrt{m} \, e^{\bbox{\iota} i \beta} \, j~,$ &
~~~~~$E>0~,~~\mbox{spin}~ \downarrow~,$\\
$\sqrt{m} \, e^{\bbox{\iota} i \beta}~\bbox{\iota}~,$ & 
~~~~~$E<0~,~~\mbox{spin}~ \uparrow~,$\\
$\sqrt{m} \, e^{\bbox{\iota} i \beta} \, \bbox{\iota} j~,$ &
~~~~~$E<0~,~~\mbox{spin}~ \downarrow~,$
\end{tabular}
\end{center}
All the imaginary units present in the complexified quaternionic Dirac theory
have now a geometric interpretation
\begin{center}
\begin{tabular}{lccl}
$\bbox{\iota}$ &~(pseudo-scalar)~ & ~~~$\rightarrow$~~~& 
positive/negative energy flip~,\\
$j$ &~(bivector)~ & ~~~$\rightarrow$~~~& 
up/down spin flip~,\\
$i$ &~(bivector)~ & ~~~$\rightarrow$~~~& 
generator of rotations~.
\end{tabular}
\end{center}


\section{From Quaternions to Clifford Algebra}
\label{s4}

In the last years, different formulations of the Dirac equation have been 
done by using noncommutative hypercomplex number. The first attempt, 
performed by using 
real quaternions~\cite{RQUA2}, gave the possibility to reduce the dimension 
of the $\gamma^{\mu}$-matrices ($4 \rightarrow 2$). The doubling of 
solutions, due to the $i$-complex geometry, allow us to obtain the standard 
results notwithstanding the halved dimension of Dirac spinors. Nevertheless, 
we also find a different representation for operators ($2 \times 2$ matrices)
and spinors (two-dimensional column vectors). The possibility to perform a
one-dimensional complexified quaternionic version of the Dirac equation by 
using $\bbox{\iota}$-complex geometry~\cite{CQUA}, overcome this problem,
by putting on the same level the operators and the spinors. Yet, it was 
not possible to give a clear geometric interpretation because of the 
complicated structure of spinors and CPT operations. The formulation of
Dirac equation by complexified quaternion and $i$-complex geometry maintains
the possibility to treat operators and spin at the same level, and more it 
permits interesting geometric interpretations.

The Dirac theory can be formulated by complex or non commutative numbers. 
The passage from 
\begin{center}
complex~~~~~$\rightarrow$~~~~~real quaternions
~~~~~$\rightarrow$~~~~~complexified quaternions
\end{center}
is achieved by performing a set of translation rules~\cite{DR1}. We can
obtain the same result by working only with the general properties of the 
Clifford algebra, in special the concept of even sub-algebra. In doing it, we 
recall the main step performed in the agreeable and clear paper of 
Zeni~\cite{ZEN}.

In the standard matrix form the Dirac equation for a free particle is usually
presented as
\begin{equation}
i \gamma^{\mu} \partial_{\mu} \psi = m \psi~,
\end{equation}
where the $\gamma^{\mu}$ matrices satisfy the relations:
\[ \gamma^{\mu} \gamma^{\nu} + \gamma^{\nu} \gamma^{\mu} = 2 g^{\mu \nu}~,
~~~~~~~\mu, \nu=0,1,2,3\]
where
\[ g^{\mu \nu} \equiv \mbox{diag} \, (+, -, -, -)\]
is the Minkowski metric. The Dirac matrices can be represented by the 
following $4\times 4$ complex matrices~\cite{BJO}
\[ \gamma^{0} = \left( \begin{array}{cc} 1 & 0 \\ 0 & $-$1 
                       \end{array} \right)~,~~~~~~~
   \vec{\gamma} = \left( \begin{array}{cc} 0 & $-$\vec{\sigma} \\ 
                         \vec{\sigma}  & 0 
                       \end{array} \right)~.
\]
A left ideal in a matrix algebra is a linear subspace of the matrix space 
which is invariant under left multiplication. The set of all matrices which 
have all elements null except those in the first column is an example of an 
ideal. The main point in the ideal approach is to put on the same level the 
operators and the spinors, representing all of them by $4\times 4$ complex 
matrices. This can be done if we use the following matrix representation
for the Dirac spinor:
\[ \Psi = \left( \begin{array}{cccc} \psi_1 & 0 & 0 & 0 \\
                                     \psi_2 & 0 & 0 & 0 \\
                                     \psi_3 & 0 & 0 & 0 \\
                                     \psi_4 & 0 & 0 & 0 
          \end{array} \right)~~~\mbox{instead of}~~~
      \psi = \left( \begin{array}{c} \psi_1 \\
                                     \psi_2 \\
                                     \psi_3 \\
                                     \psi_4  
          \end{array} \right)~.
\]
The formulation of the Dirac equation using $\psi$ or $\Psi$ are 
completely equivalent. The 
ideals of an algebra are generated by idempotents, i.e. elements of 
the algebra whose squares are equal to themselves. For example, the 
following matrix is the generator of the ideal to which $\Psi$ belongs:
\[ U = \left( \begin{array}{cccc}    1 & 0 & 0 & 0 \\
                                     0 & 0 & 0 & 0 \\
                                     0 & 0 & 0 & 0 \\
                                     0 & 0 & 0 & 0 
          \end{array} \right)~.
\]
It is clear that $U^2=U$ and also that for every $A \in C(4)$, $AU$ is an 
element of the ideal defined above (only the elements in the first column 
are non null), so that every $\Psi$ can be replaced by the product $AU$.
Thus, the Dirac equation now reads:
\begin{equation}
\gamma^{\mu} \partial_{\mu} AUi = m AU~.
\end{equation}
                           
The Clifford algebra $Cl_{4,1}$ is generated by the vectors 
$\xi_x$, $x\in [0,4]$ of $Cl_{4,1}$ that satisfy the following relation:
\[ \xi_x \xi_y + \xi_y \xi_x = 2 g_{xy}~,\]
where
\[ g^{xy} \equiv \mbox{diag} \, (-, +, +, +, +)~.\]
It is well know that the algebra $Cl_{4,1}$ is isomorphic to $C(4)$, so the
Dirac equation has a natural representation in the Clifford algebra 
$Cl_{4,1}$ in the sense that all elements present in the Dirac equation 
belong to $Cl_{4,1}$. For example, the Dirac matrices $\gamma^{\mu}$ 
are the representatives in $C(4)$ of products of vectors $\xi$. We write the
representative and the matrices with the same symbol since this does not
cause any confusion, 
\[ \gamma^{\mu} \equiv \xi^{\mu} \xi^{4}~~~~~~~\mu=0,1,2,3\]
and the imaginary unit of $C(4)$ is the volume element of $Cl_{4,1}$ 
given below:
\[ i \equiv \xi^0 \xi^1 \xi^2 \xi^3 \xi^4 ~.\]
We can see from the standard representation of Dirac matrices that the
idempotent $U$ can be written as follows:
\[ U = \frac{1}{4} \left( 1 + \gamma^0 \right) 
                   \left( 1 - i \gamma^{1} \gamma^{2} \right)~.
\]
 It is easy to verify the following identities, which will be useful 
afterwards:
\begin{eqnarray*}
\gamma^0 U = U \gamma^0 = U~,\\
iU = U i = \gamma^2 \gamma^1 U = U \gamma^2 \gamma^1~.
\end{eqnarray*}
The last equation shows to us that the imaginary unit, when multiplied by 
the idempotent, can be replaced by the product $\gamma^2 \gamma^1$. The 
Dirac equation can be represented in the Clifford algebra $Cl_{4,1}$
through the following expression:
\begin{center}
\begin{tabular}{|c|} \hline \\
~~$\gamma^{\mu} \partial_{\mu} AU \gamma^2 \gamma^1  = m AU \gamma^0$~~\\
 \\
\hline 
\end{tabular}
\end{center}
It is well known that the Clifford algebras are $Z_2$ graded, i.e., we can 
divide the linear space of the algebra in two subspaces of even and odd 
grades, which are respectively composed of even and odd products of the 
generators. The even and odd subspaces of $Cl_{4,1}$ are spanned by the 
following sets:
\[ Cl_{4,1}^{+} = \{ 1,\xi_x \xi_y, i \xi_x \}~~~~~~\mbox{and}
   ~~~~~Cl_{4,1}^{-} = \{ \xi_x,  i\xi_x \xi_y , i \}~.
\]
Every element of the odd subspace $Cl_{4,1}^{-}$ can be written as the 
product of an element of the even sub-algebra by an element of the odd 
subspace. In particular we have
\[ Cl_{4,1}^{-} = Cl_{4,1}^{+} i \gamma^2 \gamma^1~.\]
Now consider a generic element of the Dirac algebra, $A \in Cl_{4,1}$. 
According to previous equation
\[ A = A^+ + A^- = A^+ + B^+ i \gamma^2 \gamma^1~,\]
where $A^+$ and $B^+$ belong to the even sub-algebra 
$Cl_{4,1}^{+} \sim Cl_{1,3}$, while $A^-$ belongs to the odd subspace
$Cl_{4,1}^{-}$. Taking the product of $A$ by the idempotent $U$  we get:
\[ AU = (A^+ + B^+ ) U = \phi U~.\]
The final step to reduce the Dirac equation to the space-time algebra (STA)  
$Cl_{1,3}$ is to take apart the idempotent which is the only element that 
does not belong to the even sub-algebra of $Cl_{4,1}$. We note that the
idempotent $U$ is the product of two idempotents
\[ u = \frac{1}{2} \left( 1 + \gamma^0 \right)~~~~~\mbox{and}~~~~~
   v = \frac{1}{2} \left( 1 + i \gamma^1 \gamma^2  \right)~.
\]
The Dirac equation is rewritten as follows:
\[
\left( 
\gamma^{\mu} \partial_{\mu} \phi \gamma^2 \gamma^1 - m \phi \gamma_0  
\right) U
= 0 ~,
\]
where the term in brackets is clearly an element of the STA,
since $\gamma^{\mu}$ and $\phi$ belong to $Cl_{1,3}$. So the above equation
looks like:
\[ Cl_{4,1}^+ U =  Cl_{4,1}^+ uv =  Cl_{4,1}^+ v ~.\] 
The idempotent $v$ is irrelevant to the above equation
\[ Dv=0~~~~~\rightarrow~~~~~D=0~~~\mbox{and}~~~D i\gamma^1 \gamma^2 = 0~,\]
and because $i \gamma^1 \gamma^2 $ is an invertible element, we have simply
$D=0$. In conclusion, the Dirac equation can then be written using only 
elements of the STA:
\begin{center}
\begin{tabular}{|c|} \hline \\
~~$\gamma^{\mu} \partial_{\mu} \phi \gamma^2 \gamma^1 u = 
m \phi \gamma_0 u$~~\\
 \\
\hline 
\end{tabular}
\end{center}

Before going on we observe that the idempotent $u$ is irrelevant in the above
equation. All mathematical and physical informations contained in 
$\phi= \phi(x) \in Cl_{1,3}$, which will be shown to be a sum of inhomogeneous 
even multivectors of $Cl_{1,3}$, and which looks as a superfield~\cite{SUP}. 
To reduce the Dirac equation to the Pauli algebra, $Cl_{3,0}$, we use the fact
that the Pauli algebra is the even sub-algebra of STA. The first step is 
analogous to the previous case
\[ \phi = \phi^+ + \phi^- = \phi^+ + \varphi^+ \gamma^0~,\]
and consequently we have
\[ \phi u = ( \phi^+ + \varphi^+ ) u = \eta u~,\]
where $u=\frac{1}{2} (1+\gamma^0)$ is the idempotent present in the  
Dirac equation performed by STA 
and $\eta$ is an element of the Pauli algebra. Multiplying the Dirac equation  
by $\gamma^0$ on the left and using $\eta$ instead of 
$\phi$, the Dirac equation becomes:
\[
\left( \gamma^0 \gamma^{\mu} \partial_{\mu} \eta \gamma^2 \gamma^1 - 
m \gamma^0 \eta \gamma_0 \right) u = 0~.
\]
The term in brackets belongs to the Pauli algebra, $Cl_{3,0} \sim Cl_{1,3}^+$,
since $\eta$ and the products $ \gamma^0 \gamma^{\mu}$ and 
$ \gamma^2 \gamma^{1}$ belong to  $Cl_{3,0}$. For the final step we have 
only to consider that
\[ Du=0~~~~~\rightarrow~~~~~D=0~~~\mbox{and}~~~D\gamma^0 = 0~,\]
and because $\gamma^0 $ is an invertible element, we have simply
\begin{center}
\begin{tabular}{||c||} \hline \hline \\
~~$\left( \partial_0 + \vec{\sigma} \cdot \vec{\partial}\right) 
\eta i \sigma_3 = m \eta^{\bullet}$~~\\
 \\
\hline \hline
\end{tabular}
\end{center}
This equation contains only element of the Pauli algebra. 
By recalling the isomorphism between the Pauli algebra and complexified 
quaternions
\begin{eqnarray*}
\vec{\sigma} & ~~\leftrightarrow ~~ & \bbox{\iota} \vec{h} ~,\\
i \vec{\sigma} & ~~\leftrightarrow ~~ & \vec{h} ~,
\end{eqnarray*}
we re-obtain the complexified quaternionic version of the Dirac equation. We 
only note that to have the desired geometric interpretations is important 
to adopt $i$-complex geometry in defining scalar products.


\section{Conclusions}
\label{s5}

The main goal of this paper is to get geometric interpretations in the 
formulation of Dirac theory by complexified quaternions and $i$-complex 
geometry. The possibility to write down a one-dimensional version of 
Dirac equation  simplify the solution of this equation. In the 
complexified quaternionic algebra we work in eight-dimensional space over 
the real numbers, while in the standard (complex) formulation we have to 
do with a 32-dimensional space on the reals. 
Moreover, all the elements in complexified quaternionic numbers have a 
geometric interpretation and this allow a clear interpretation for the 
complexified quaternionic imaginary units
\[ \bbox{\iota}~,~~~$i$~,~~~$j$~,\]
which appears in the spinor structure. We also recall that the Dirac 
spinors assume in  the complexified quaternionic {\em polar} representation 
a very simple form. For these reasons, we think that the use of the 
complexified quaternionic algebra (together $i$-complex geometry) 
represents the ``natural'' mathematical
language to write down the Dirac equation and 
discuss electron theory.


\acknowledgements

One of the authors (SdL) is indebted to Edmundo, Z\'e Emilio, Jayme,
Mar\c{c}ia, S\^onia, Cristiane and Heloiza for their friendly and warm 
hospitality during the stay in Brasil. 
The author also gratefully acknowledges the 
IMECC-UNICAMP for financial support.


\end{document}